\documentclass[11pt]{article}
\pdfoutput=1
\usepackage{jcapmod}
\usepackage{booktabs}
\usepackage[english]{babel}
\usepackage{amsmath, amssymb, amsfonts, amsbsy, amstext, amsthm}
\usepackage{graphicx}
\usepackage{exscale}
\usepackage[makeroom]{cancel}
\usepackage{soul}
\usepackage{mathtools}
\usepackage{multirow}

\allowdisplaybreaks[1]

\numberwithin{equation}{section}

\def\beq{\begin{equation}}
\def\eeq{\end{equation}}
\def\bea{\begin{eqnarray}}
\def\eea{\end{eqnarray}}

\def\k{{\boldsymbol{k}}}

\def\x{\vec{x}}

\def\k{{\bf k}}

\def\x{{\bf x}}

\def\fNL{f_{\rm NL}}

\def\fNL{f_{\rm NL}}

\def\za{{\zeta_{\rm ad}}}
\def\zaL{{\zeta_{{\rm ad}, L}}}
\def\zaS{{\zeta_{{\rm ad}, S}}}
\DeclareRobustCommand{\SkipTocEntry}[4]{}

\begin{document}

\pagenumbering{roman}
\begin{titlepage}
\baselineskip=15.5pt \thispagestyle{empty}

\bigskip\

\vspace{1cm}
\begin{center}

{\fontsize{20.74}{24}\selectfont \sffamily \bfseries  Single-Field Inflation and the Local Ansatz:\\
Distinguishability and Consistency}

\end{center}
\vspace{0.2cm}
\begin{center}
{\fontsize{12}{30}\selectfont Roland de Putter,$^{\blacktriangledown}$ Olivier Dor\'e,$^{\bigstar,\blacktriangledown}$, Daniel Green,$^{\spadesuit,\clubsuit}$ and Joel Meyers$^{\clubsuit}$}
\end{center}

\begin{center}

\vskip 8pt
\textsl{$^\blacktriangledown$ California Institute of Technology, Pasadena, CA 91125}
\vskip 7pt

\textsl{$^\bigstar$ Jet Propulsion Laboratory, California Institute of Technology, Pasadena, CA 91109}
\vskip 7pt

\textsl{$^\spadesuit$ University of California, Berkeley, California 94720, USA}
\vskip 7pt

\textsl{$^\clubsuit$ Canadian Institute for Theoretical Astrophysics, Toronto, ON M5S 3H8, Canada}
\vskip 7pt

\end{center}

\vspace{1.2cm}
\hrule \vspace{0.3cm}
\noindent {\sffamily \bfseries Abstract} \\[0.1cm]
The single-field consistency conditions and the local ansatz have played separate but important roles in characterizing the non-Gaussian signatures of single- and multifield inflation respectively.  We explore the precise relationship between these two approaches and their predictions.  We demonstrate that the predictions of the single-field consistency conditions can never be satisfied by a general local ansatz with deviations necessarily arising at order $(n_s-1)^2$.  This implies that there is, in principle, a minimum difference between single- and (fully local) multifield inflation in observables sensitive to the squeezed limit such as scale-dependent halo bias.  We also explore some potential observational implications of the consistency conditions and its relationship to the local ansatz. In particular, we propose a new scheme to test the consistency relations. In analogy with delensing of the cosmic microwave background, one can deproject the coupling of the long wavelength modes with the short wavelength modes and test for residual anomalous coupling.

\vskip 10pt
\hrule
\vskip 10pt

\vspace{0.6cm}
\end{titlepage}

\thispagestyle{empty}
\setcounter{page}{2}
\tableofcontents

\clearpage
\pagenumbering{arabic}
\setcounter{page}{1}

\section{Introduction}
\label{sec:intro}

Understanding the origin of the initial conditions for the universe is one of the primary goals of modern cosmology.  Most ambitiously, we hope to test fundamental principles behind the origin of structure, independently of any framework.  For example, one might hope to distinguish whether the initial seeds are the result of quantum or classical fluctuations~\cite{Maldacena:2015bha}.  Even within the context of inflation, we would like to test the nature of inflation, including whether inflation was single- or multifield~\cite{Creminelli:2004yq} or if inflation is a weakly or strongly coupled phenomenon~\cite{Baumann:2014cja,Baumann:2015nta}.  Significant progress has been made in identifying possible observational targets~\cite{Alvarez:2014vva}, often in the context of deviations from Gaussianity.  Still, many of these targets are qualitative in nature and more work remains to connect them to fundamental principles~\cite{Baumann:2015nta}.

Perhaps the most quantitative tools for testing inflation are the single-field consistency conditions~\cite{Maldacena:2002vr,Creminelli:2004yq}.  They state that when inflation is driven by a single field (or clock), the coupling of short and long modes is completely specified: ($N$ +1)-point correlation functions involving short and long modes can be specified in terms of lower order correlation functions. These relations are testable observationally.

The basic reason underlying these conditions is that, to leading order in gradients, the long mode metric fluctuation $\zeta_L$ is locally a constant that is equivalent to a re-parameterization of the clock.  This logic has been extended to show the long mode has no {\it local} physical effects up to quadratic order in gradients~\cite{Creminelli:2004yq,Baldauf:2011bh,Creminelli:2012ed}. As such, the statement of the consistency conditions is essentially that, modulo gradients of the long mode, the short modes cannot measure the presence of the long mode physically. The leading order effect of the long mode that {\it can} be measured locally is a perturbation to the local curvature, which is suppressed by $k_L^2$, where $k_L$ is the wave-number of $\zeta_L$.

Whereas these consistency conditions were initially introduced by Maldacena to explain the properties of inflationary correlation functions~\cite{Maldacena:2002vr}, they have since been found to have very general consequences to cosmology~\cite{Creminelli:2004yq}, even at much later times.  The essence of these consistency conditions was understood much earlier in the context of the separate universe approach (see e.g.~\cite{PhysRevD.31.1792,Salopek:1990jq}).  Weinberg~\cite{Weinberg:2003sw} later understood that these are all consequences of a large gauge transformation that may be implemented at any time (not just during inflation), which has ultimately made a number of powerful applications possible.  In particular, it was shown to be straightforward to predict the implications of the consistency conditions for any observable and thus look for deviations~\cite{Creminelli:2013mca,dePutter:2015vga,Dai:2015jaa}.

Since the full set of consistency conditions strongly constrains the statistics of the initial conditions, it is natural to compare these constraints to those stemming from a common prescription for the initial conditions, namely the local ansatz.  The local ansatz simply assumes that there exists some Gaussian random field $\zeta_g(\x,t)$ such that the initial conditions for the adiabatic mode are generated {\it locally} in this Gaussian field:
\beq
\zeta(\x,t_i ) = \sum_n c_n \zeta_g(\x,t_i)^n = \zeta_g + \frac{3}{5} \fNL^{\rm local} \zeta_g^2 + \ldots \ .
\eeq
Data from the {\it Planck} satellite currently constrain $\fNL^{\rm local} = 0.8 \pm 5.0$~\cite{Ade:2015ava} but future observations have the potential to reach $\sigma(\fNL)<1$ \cite{RdPDore14,SPHERExWP,ferrsmith14,yamauchietal14}.  This is particularly interesting as $|\fNL| > 1$ is a common feature of models that reproduce the local ansatz~\cite{lythetal03,zal04,Alvarez:2014vva}.

The idea that some non-linear but local physics generated the initial conditions is very plausible and is indeed found to arise in many multifield models of inflation and alternatives to inflation.  Nevertheless, the origin of the local ansatz in physical examples is qualitatively different from the single-field consistency conditions.  The local ansatz is usually the consequence of local non-linear evolution at times when all the observable modes are outside the horizon.  Since there are no physical scales larger than the horizon, long and short wavelength modes are treated on the same footing.  While local interactions also govern the single-field consistency conditions, only the long wavelength modes are outside the horizon and therefore the short and long modes are physically distinguishable in the resulting statistics.

Given the differences in the physics, it is natural to ask at what level one expects to find deviations in predictions made by the local ansatz and single-field inflation.  This is particularly important when testing observationally the nature of inflation.
The statement that the consistency conditions imply that $f_{NL} = -{5\over 12} (n_s - 1)$ would seem to suggest that single-field inflation is equivalent to a local ansatz with specific coefficients. As we will show explicitly, this statement is not correct. First of all, the single-field consistency conditions are really an infinite set of constraints rather than just a statement of a single statistic~\cite{Li:2008gg} and matching the above relation would only confirm one from this infinite set. Second, as will be discussed further below,  these relation between $f_{NL}$ and $n_s$ involves statistically average quantities whereas the consistency conditions should hold for any realization and not just statistically. This suggests that mapping the single-field consistency conditions onto parameters predicted by the local ansatz mischaracterizes the relevant physical effects.

Another motivation for this work is to further clarify the observability of the single-field consistency conditions.  As has been emphasized by a number of authors, the consistency conditions physically imply that the short modes are statistically independent of the long mode, in physical coordinates.  In this sense, single-field inflation predicts ``zero mode-coupling" which suggests there is no natural target for local non-Gaussianity, even in principle~\cite{Senatore:2012wy,Pajer:2013ana,Dai:2015jaa}.  Nevertheless, as we will show, the local ansatz can never reproduce this prediction; it leaves a non-zero mode coupling at least of order $(n_s-1)^2$ in any such observable and therefore sets a natural target (although unobservable in practice).  For example, the local ansatz will always lead to scale-dependent bias\footnote{Here, scale-dependent bias refers to any term in the bias expansion which is not consistent with locality in space.  This includes terms like $\zeta_L^{n>2}$ which are non-local and also non-linear.} while single-field inflation does not~\cite{Pajer:2013ana,dePutter:2015vga,Dai:2015jaa}.

In this paper, we will explore the relationship between the consistency conditions and the local ansatz.  In Section~\ref{sec:violation}, we will show that the local ansatz cannot reproduce the consistency conditions for any choice of parameters.  In Section~\ref{sec:consistency}, we describe how the local ansatz needs to be modified to be consistent with Weinberg's derivation of the consistency conditions.  In Section~\ref{sec:deprojecting}, we will demonstrate how the mode coupling induced in single-field inflation can be deprojected from the observed statistics in direct analogy with weak lensing of the cosmic microwave background.

\section{Violating the Single-Field Consistency Conditions}
\label{sec:violation}

In this section, we will show that the local ansatz, $\zeta(\x) = \sum_n c_n \zeta_g(\x)^n$, cannot satisfy the single-field consistency conditions for any choice of $c_n$.  It will be important that the coefficients $c_n$ cannot depend on the location in space because we are assuming that only $\zeta_g(\x)$ breaks homogeneity. Therefore, $c_n$ is a list of numbers rather than functions. 

The qualitative reason these two models do not agree can be understood as follows.  The local ansatz cannot distinguish long and short modes (as required by locality), and therefore a given coefficient predicts that a number of different mode couplings are related.  This is particularly important for $c_{n>2}$ as there is more than one long-short coupling per coefficient.  If the local ansatz is to match the single-field consistency conditions, these non-trivial relations must also arise in single-field inflation.  However, single-field inflation distinguishes long and short modes and there is no reason to expect the same relations to hold.  The essence of this section is check that this expected difference cannot be eliminated by carefully choosing the coefficients of the local expansion.

We first need to be clear about how the consistency conditions act on correlation functions of short modes\footnote{The exact separation between short and long modes is not always precise.  Very conservatively, requiring $k_L/ k_S < {\cal O}(100)$ should guarantee that we are in the squeezed limit where the consistency conditions apply~\cite{Flauger:2013hra}.  For many single-field models, a small hierarchy is sufficient.}.  Let us start with a metric without a long mode such that
\beq
d\tilde s^2 = -dt^2 + a(t)^2 \, e^{2 \tilde\zeta_S(\tilde \x)} \, d\tilde \x^2  \ .
\eeq
 Now we introduce the long mode through the transformation $\x = e^{-\zeta_L} \tilde \x$, which implies
\beq
ds^2 = d\tilde s^2 = -dt^2 + a(t)^2 \, e^{2 \tilde \zeta_S(\tilde \x)} \, d\tilde \x^2= -dt^2 + a^2(t) \, e^{2 \tilde \zeta_S(e^{\zeta_L} \x)+ 2 \zeta_L} \, d \x^2 \ .
\eeq
where $\tilde \zeta_S$ is the original short perturbation that is independent of $\zeta_L$.  Throughout, we will ignore all gradients\footnote{We can extend these results to linear order in gradients using the conformal consistency conditions~\cite{Creminelli:2012ed}.} of $\zeta_L$ and keep only the leading order behavior in the limit of vanishing wave-number, $k_L \to 0$.  The resulting transformation of the short mode is
\beq
\label{eq:c.c. def}
\zeta_S(\x) = \tilde{\zeta}_S (e^{\zeta_L} \x) = \tilde{\zeta}_S(\x) + \zeta_L \, \x \cdot \nabla \tilde{\zeta}_S(\x) + \ldots \ .
 \eeq
Thus, in the presence of a long mode $\zeta_L$, all ``local'' statistics of $\zeta$, i.e.~$N-$point functions purely of the short modes $\zeta_S$, can be obtained
by evaluating the same quantities in the absence of the long mode, but at a different scale, $\x \to \x \, e^{\zeta_L}$, or $\k \to \k \, e^{-\zeta_L}$.

We will compare this to the local ansatz, which we will define as
\beq
\label{eq:def local}
\zeta(\x,t_i ) = \sum_n c_n \zeta_g(\x,t_i)^n = \zeta_g + \frac{3}{5} \fNL^{\rm local} \zeta_g^2 + \ldots
\eeq
where from here on, we will drop the dependence on the initial time $t_i$.
Here $\zeta_g$ is assumed to satisfy Gaussian statistics and therefore $\zeta_{g,S}$ and $\zeta_{g,L}$ are statistically independent\footnote{The perturbation $\tilde{\zeta}_S$ appearing in the consistency conditions is simply the small-scale perturbation in the absence of the long mode so that
$\tilde \zeta_S$ is not in general equal to $\zeta_g$ because we have made no assumption about the statistics of $\tilde{\zeta}_S$.}.  The local ansatz thus leads to a mode coupling,
\bea
\label{eq:local mc}
\zeta_S &=& c_1 \, \zeta_{g,S} + c_2 \, \zeta_{g,S}^2 + c_3 \, \zeta_{g,S}^3 +  \ldots \nonumber \\
&+& \zeta_{g,L} \, \left( 2 \, c_2 \, \zeta_{g,S} + 3 \, c_3 \, \zeta_{g,S}^2 + \ldots \right) \nonumber \\
&+& \zeta_{g,L}^2 \, \left( 3 \, c_3 \, \zeta_{g,S} + \ldots  \right).
\eea
While this series extends to arbitrary orders in $\zeta_{g,S}$ and $\zeta_{g,L}$ it is important that the modulation of a connected $(N+1)$-point correlation function of short modes at ${\cal O}\left(\zeta_{g, L}^q \right)$ is determined by $c_{n \leq N+q}$ (ignoring loop-suppressed corrections).

Now, let us examine what the single-field consistency conditions predict for the behavior of the two-point statistics of the short modes.  Up to second order in the long mode, we find
\begin{align}
\label{eq:cc all orders1}
	\langle \zeta_S({\bf k} ) \, \zeta_S({\bf k}' ) \rangle' &= \langle \tilde{\zeta}_S({\bf k} \, e^{- \zeta_L}) \, \tilde{\zeta}_S({\bf k}' \, e^{- \zeta_L}) \rangle' = e^{-(n_s-1) \zeta_L} \, \langle \tilde{\zeta}_S({\bf k}) \, \tilde{\zeta}_S({\bf k}') \rangle' \nonumber \\
	&= P_S(k) - (n_s-1) \, P_S(k) \, \zeta_L + \frac{1}{2}(n_s-1)^2 \, P_S(k) \, \zeta_L^2 + \mathcal{O}(\zeta_L^3) \, .
\end{align}
The primes in the first line indicate that we suppress the usual factor $(2 \pi)^{3} \, \delta^{(D)}(\sum_i {\bf k}_i)$ from the expectation value.
We have taken $n_s$ to be constant, since if it depended on scale, the local ansatz would fail to match the predictions of the single-field consistency conditions. Thus, a first (trivial) requirement for the local ansatz to reproduce the predictions of the consistency conditions is that the spectral index is scale-independent.

Up to second order in the long mode, we find for the local ansatz
\begin{align}
\label{eq:local_power}
	\langle \zeta_S({\bf k} ) \, \zeta_S({\bf k}' ) \rangle' &=\left(c_1^2 + 4c_1c_2 \, \zeta_L + (4c_2^2 + 6c_1c_3) \, \zeta_L^2\right)\langle \zeta_{g,S}({\bf k} ) \, \zeta_{g,S}({\bf k}' ) \rangle' + \mathcal{O}(\zeta_L^3) \nonumber \\
	&=c_1^2P_S(k) + 4c_1c_2P_S(k) \, \zeta_L + (4c_2^2+6c_1c_3)P_S(k) \, \zeta_L^2 + \mathcal{O}(\zeta_L^3) \, .
\end{align}
Matching the two-point predictions of the single-field consistency conditions order by order requires that we have $c_1=1$, $c_2=-\frac{1}{4}(n_s-1)$ (i.e.~the familiar $\fNL = - \frac{5}{12} (n_s - 1)$), and $c_3=\frac{1}{24}(n_s-1)^2$.  This choice of coefficients then dictates the behavior of the three-point function of short modes for the local ansatz
\begin{align}
\label{eq:local_bispectrum}
	\langle \zeta_S \zeta_S \zeta_S \rangle' &= 6c_1^2c_2P_S(k)^2 + (18c_1^2c_3 + 24c_1c_2^2)P_S(k)^2\zeta_L + \mathcal{O}(\zeta_L^2) \nonumber \\
	&= - \frac{3}{2} (n_s-1)P_S(k)^2 +  \frac{9}{4} (n_s-1)^2 P_S(k)^2\zeta_L + \mathcal{O}(\zeta_L^2) \, .
\end{align}

Returning to the predictions of single-field inflation, we are free to choose the form of the bispectrum containing only short modes since that correlation is unconstrained by symmetries (although it would be very constraining if the only way to reconcile the local ansatz with single-field inflation is for this exact form of the local bispectrum).  Once this choice is made, however, the scaling of the bispectrum with long modes is completely determined by the single-field consistency conditions
\begin{align}
\label{eq:cc_bispectrum}
	\langle \zeta_S \zeta_S \zeta_S \rangle' &= - \frac{3}{2} (n_s-1)P_S(k)^2 e^{-2 (n_s-1) \zeta_L} \nonumber \\
	&= - \frac{3}{2} (n_s-1)P_S(k)^2 +  3 (n_s-1)^2 P_S(k)^2\zeta_L + \mathcal{O}(\zeta_L^2) \, .
\end{align}
Comparing Eqs.~(\ref{eq:local_bispectrum}) and (\ref{eq:cc_bispectrum}), we see that if the coefficients of the local ansatz are chosen to make the behavior of the two-point statistics of the short modes match the predictions of the consistency conditions, then the predictions for the bispectrum necessarily disagree at $\mathcal{O}\left((n_s-1)^2\right)$. Furthermore, we cannot correct this disagreement by introducing additional terms to the local ansatz with $c_{n>3}$ because no such terms contribute to the three-point statistics of the short modes at first order in $\zeta_L$ (except through loops which are highly suppressed).

The origin of this contradiction can be generalized to arbitrary orders in $\zeta_L$.  Suppose we truncate the local expansion at order $\zeta^N$.  In this case, once we make the split into long and short modes, we have
\beq
\zeta = \sum_{n=1}^{N} c_n (\zeta_{g,S} + \zeta_{g,L})^n
\eeq
we can always fix $c_1=1$ by definition.  This means we have $N-1$ unknown coefficients to match $\langle \zeta_{g,S}^m \rangle$ to order $\zeta_L^{N-m+1}$ where $m = 2...N$.  We find that there are $\sum_{m=2}^{N} (N-m+1) =\sum_{i=1}^{N-1} i =N\times (N-1)/2$ different coefficients that we need to match using these $N-1$ unknown coefficients.  This system is therefore overconstrained and it would thus be a miracle if the coefficients matched the consistency conditions.
\vskip 10pt
We can see that the general pattern matches the explicit calculations including $c_{1,2,3}$.  For $N=2$, we have one coefficient ($c_2$) but we only have to match one number, the squeezed limit of the bispectrum.  At order $N=3$, we have 2 coefficients $c_2, c_3$ but now we have 3 different squeezed limits to match and we simply cannot pick $c_2$ and $c_3$ to make them all agree with the single-field consistency conditions.  At order $N$ we should find that floor($N/2$) consistency conditions cannot be satisfied by the local ansatz.

\vskip 10pt
\noindent {\it Summary:} \hskip 6pt We have shown that it is impossible to exactly obey the single-field consistency conditions with the local ansatz. In that sense, testing the single- vs.~multifield nature of inflation by constraining $f_{\rm NL}$, etc., within the local ansatz is technically not correct, as no point in this parameter space is consistent with single-field inflation. However, the local ansatz is of course still very useful as a shorthand description for the squeezed limit behavior of the bispectrum and/or the collapsed limit trispectrum. These are also the quantities that determine the leading order signal of scale-dependent halo bias \cite{dalaletal08,matver08,slosaretal08,desjacjak10}, which is one of the main ways in the near future to constrain primordial non-Gaussianity using large-scale structure~\cite{Alvarez:2014vva}. This is how the local ansatz is most commonly used, and in this sense the single-field case is indeed equivalent to $f_{\rm NL} = - \frac{5}{12} (n_s  - 1)$. However, if one were to use the local form to also predict e.g.~the modulation of the short-scale bispectrum, $\langle \zeta_L \, \zeta_S^3 \rangle$, and higher order modulations in $\zeta_L^2$ such as $\langle \zeta_L^2 \, \zeta_S^2 \rangle$, we have shown that one would inevitably make predictions inconsistent with single-field inflation.  Of course, in practice, these deviations from the predictions of single-field inflation are too small to be detected with any near-term observations.

\section{Consistency Conditions for the Local Ansatz}
\label{sec:consistency}

In the previous section, we found that the local ansatz can never match the predictions of the single-field consistency conditions.  Physics is rarely discontinuous and therefore we expect that there is some generalization of the local ansatz that should allow us to interpolate between the two.  This is also obvious from the point of view of model building, as we can certainly write models of inflation that interpolate between single- and multifield by varying the mass of the additional fields.  However, if we take the local ansatz as our starting point, we want to know the minimal set of terms needed to reproduce both limits.

There are two generalizations of the local ansatz that could plausibly change our results: (1) multiple random fields and (2) ``non-local" terms\footnote{We remind the reader that  {\it local} is taken in the sense of the local ansatz, i.e.~functions of the form $\Phi(\x) = F(\{\phi_i(\x)\})$.  Non-local terms need not imply a violation of causality/locality in the dynamics of $\phi$.  Non-local terms can arise when statistics have memory of past evolution and/or when there is a scale, such as the horizon, that can distinguish the wavelengths of $\phi(\x)$ (the local form necessarily treats all wavelengths on the same footing).} in the expansion in the Gaussian random field(s).  Given that the local ansatz is a prediction of multifield inflation, adding more random fields is an obvious choice.  We will see that adding multiple fields is not a sufficient condition, but that both non-local terms and multiple fields are needed to interpolate between the consistency conditions and the local ansatz.

Let us consider a scenario with perturbations in two directions (this can be straightforwardly generalized to the case of more than two fields), $\zeta$ and $\sigma$, and let us assume that any shift in the perturbation with $\Delta \sigma = 0$ implies the shift is along the adiabatic direction. Varying $\sigma$ at $\zeta=0$ then of course describes an isocurvature fluctuation\footnote{As a simple example, in the case with two scalar fields $\phi = \bar{\phi} + \delta \phi$ and $\chi = \bar{\chi} + \delta \chi$,
a commonly considered scenario is one where the curvature-isocurvature basis is approximately aligned with the $\delta \phi$-$\delta \chi$ basis,
so that $\zeta \approx - \frac{H}{\dot{\bar{\phi}}} \, \delta \phi$, and $\sigma \approx \delta \chi$.
This is typically the case for the initial conditions in models where $\chi$ is a spectator field during inflation.}.

The single-field consistency conditions in this more general context are really consistency conditions about the effects of an {\it adiabatic} shift in the long-mode fluctuation (see e.g.~\cite{Senatore:2012wy,LopezNacir:2012rm, Goldberger:2013rsa,Creminelli:2013mca,Mirbabayi:2015hva} for related discussions). Specifically, the generalization of the single-field consistency conditions, Eq.~(\ref{eq:c.c. def}), is that under such a transformation,
\bea
\label{eq:c.c. multi def}
\tilde{\zeta}(\x) &\to&  \zeta(\x) =  \tilde{\zeta}_S(e^{\Delta \zeta_L} \, \x ) + \tilde{\zeta}_L + \Delta \zeta_L \nonumber \\
\tilde{\sigma}(\x) &\to& \sigma(\x) = \tilde{\sigma}_S(e^{\Delta \zeta_L} \, \x ) + \tilde{\sigma}_L \, ,
\eea
where quantities with a tilde are the fields in the absence of the shift $\Delta \zeta_L$, which must be statistically independent of $\Delta \zeta_L$.
If all we wanted was to express the consistency conditions in a multifield scenario, Eq.~(\ref{eq:c.c. multi def}) would be sufficient. However, the above expression does not fully specify the statistics of the curvature perturbation (nor of $\sigma$), as it does not say anything about the statistics of $\tilde{\zeta}_S$ and $\tilde{\sigma}$, other than their independence of $\Delta \zeta_L$. In particular, we have not fully specified the response of $\zeta_S$ to long modes, because we have not specified the response to $\sigma_L$.

The usual {\it local ansatz}, Eq.~(\ref{eq:def local}), fixes the {\it full} statistics of the curvature perturbation by expressing $\zeta$ as a local function of a {\it Gaussian} field $\zeta_g$. We would like to do the same here, but using the presence of $\sigma$ (or in general of multiple fields) to remain in agreement with the consistency conditions. Specifically, we would like to express the perturbations in terms of two
Gaussian fields, $\za$ (an adiabatic fluctuation) and $\sigma_g$, to fully specify the statistics\footnote{The fields technically do not have to be Gaussian. To specify the mode coupling, we really only need to demand that the short-mode components of $\za$ and $\sigma_g$ are independent of the long-mode components.}.
Based on Eq.~(\ref{eq:c.c. multi def}), a minimal consistent ansatz we could write is,
\bea
\zeta(\x) &=& \za_{,S}(e^{\za_{,L}} \, \x ) + \za_{,L} \nonumber \\
\sigma(\x) &=& \sigma_g(e^{\za_{,L}} \, \x ) \, ,
\eea
with the component fields Gaussian.
Now, the most general\footnote{It is straightforward to check that any other local term is not allowed in Eq.~(\ref{eq:local trans}).  The contributions $\zeta \to F(\zeta) = F(\za_{,S}(e^{\za_{,L}} \, \x) + \za_{,L})$ or $\sigma \to G(\zeta) = G(\za_{,S}(e^{\za_{,L}} \, \x) + \za_{,L})$ will not obey the transformation in Equation~(\ref{eq:c.c. multi def}) unless $F(x) = x$ and $G(x) = 0$.} local transformation of this ansatz that still respects Eq.~(\ref{eq:c.c. multi def}) is,
\bea
\label{eq:local trans}
\zeta &\to & \zeta = \zeta + f(\sigma) \nonumber\\
\sigma &\to & \sigma =  g(\sigma) \, ,
\eea
leading to the generalized local ansatz,
\bea
\zeta(\x) &=& \left[ \za_{,S} + f({\sigma}_g )\right](e^{\za_{,L}} \, \x) + \za_{,L} \nonumber\\
\sigma(\x) &=& \left[g({\sigma}_g)\right](e^{\za_{,L}} \, \x) \, .
\eea
Finally, expressing the generalized local ansatz for $\zeta$ to second order, and separating short and long modes, gives,
\bea
\zeta &=&  \left[ \za_{,S} +  \sigma_{g,S} \right](e^{\za_{,L}} \, \x) + c_2 \, \sigma_{g}^2 + \za_{,L} + \sigma_{g,L} +\dots \nonumber \\
&=& \za + \sigma_g + \za_{,L} \, \x \cdot \nabla \left[ \za_{,S} +  \sigma_{g,S} \right] + c_2 \, \sigma_{g}^2 + \dots \quad  \text{{\bf (generalized local ansatz)}} \label{eq:multifield local}
\eea
where we have Taylor expanded $f$ in powers of $\sigma_g$ and then absorbed the coefficients $\partial_\sigma f$ and $\partial^2_\sigma f$ into $\sigma_g$ and $c_2$.
The statistics of $\zeta$, and in particular the mode coupling, are now fully determined by Eq.~(\ref{eq:multifield local}) as soon as the variance of $\za$ and $\sigma_g$
are specified. We choose them to be uncorrelated\footnote{If Eq.~(\ref{eq:multifield local}) holds, but $\za$ and $\sigma_g$ are a priori {\it not} independent, we can always apply a redefinition $\za \to \za' \equiv \za + A \, \sigma_g$ such that $\za'$ and $\sigma_g$ {\it are} independent. However, after the redefinition, the mode-coupling would have a slightly more general form (dropping the prime and reabsorbing some coefficients into $\sigma_g$ and $c_2$),\\
$\zeta = \left[ \za_{,S} +  \sigma_{g,S} \right](e^{\za_{,L} - \alpha \, \sigma_{g,L}} \, \x) + c_2 \, \sigma_{g}^2 + \za_{,L} + \sigma_{g,L} + \dots
$} $\langle \za \, \sigma_g \rangle = 0$.
Clearly, the restriction placed on Eq.~(\ref{eq:local trans}) by the adiabatic consistency conditions means our final form can only have significant local-type non-Gaussianity due to the presence of the second field, $\sigma_g$.

Finally, in cases where there is more than one non-adiabatic mode (more than two fields),
one can without loss of generality define $\sigma_g \equiv \sigma_{g,1}$ to be the linear combination contributing linearly to $\zeta$,
generalizing Eq.~(\ref{eq:multifield local}) so that only the quadratic term is modified,
\beq
c_2 \, \sigma_g^2  \to  \sum_{ij} c_{2,ij} \, \sigma_{g,i} \, \sigma_{g,j} \, ,
\eeq
where the sum is over all non-adiabatic modes $\sigma_{g,i}$, with $\langle \sigma_{g,i} \, \sigma_{g,j} \rangle = 0$ for $i \ne j$. The modes $\sigma_{g,i}$ with $i>1$ exclusively contribute to stochastic non-Gaussianity
because they are by definition uncorrelated with $\zeta$ at linear order.

\vskip 10pt

Equation (\ref{eq:multifield local}) is {\it not} intended to be the most general form for non-Gaussianity in multifield inflation. It is merely an ansatz that, loosely speaking, minimally satisfies the consistency conditions, and allows for all local (in the sense discussed in the beginning of this section) terms that do not violate them.  However, we have not addressed how this ansatz can arise physically.  There are two implicit assumptions about the dynamics that are crucial:
\begin{itemize}
\item The mode coupling at horizon crossing is trivial.  The horizon sets a natural scale that allows for terms that are not of the local form.  Most significantly, this would allow for terms of the form $(\zeta_S)^m \, (\sigma_L)^n$ that are allowed by the consistency conditions.  
\item The fluctuations in $\zeta$ at constant $\sigma$ correspond to the adiabatic mode that is constant in time outside of the horizon. It is this mode that can be removed by a coordinate transformation.  This is an assumption about having reached the inflationary attractor solution.  
\end{itemize}
We can make these points more concrete by considering a simple multifield inflation scenario.  The discussion below closely resembles the ``derivation'' above of the generalized ansatz.
We can decompose field perturbations in terms of curvature and isocurvature fluctuations. For instance, in a 2-field model with separable potential $W(\phi, \chi) = U(\phi) + V(\chi)$, and assuming slow-roll for simplicity, the curvature perturbation is, to first order,
\beq
\zeta = \frac{W \, U_\phi}{U_\phi^2 + V_\chi^2} \, \delta \phi + \frac{W \, V_\chi}{U_\phi^2 + V_\chi^2} \, \delta \chi + {\cal O}(\delta \phi^2, \delta\chi \delta\phi, \delta \chi^2) \ .
\eeq
While we have only included the linear order terms, $\zeta$ is defined to all orders in the fluctuations.  We can choose
\beq
\sigma \propto \frac{\delta \phi}{U_\phi} - \frac{\delta \chi}{V_\chi} \ ,
\eeq
so that $\sigma = 0$ corresponds to an adiabatic fluctuation (to first order).

Now consider initial conditions at some time when all modes of interest have just exited the horizon, indicated by a $*$ subscript.
In scenarios with two light fields, $\delta \phi_*$ and $\delta \chi_*$ are typically close to Gaussian and independent.
Writing only the minimal mode coupling required to satisfy the consistency conditions, we can then express the initial fluctuations in the $\zeta - \sigma$ basis in terms of truly independent Gaussian fields (which we will again write as $\za$ and $\sigma_g$) as,
\bea
\label{eq:initial NG}
\zeta_* &=& \za_{,S}(e^{\za_{,L}} \, \x) + \za_{,L}  \\
\sigma_* &=& \sigma_{g,S}(e^{\za_{,L}} \, \x) + \sigma_{g,L} \ . \label{eq:sigmastar}
\eea
In essence, we are assuming that the physics of horizon crossing is trivial (in local coordinates) and all subsequent evolution can be treated classically from these initial conditions\footnote{We could even allow for significant initial non-Gaussianity in $\sigma$ by adding a term ${\cal O}(\sigma_g^2)$ to Equation~(\ref{eq:sigmastar}). This would leave the final form of the statistics unchanged.  In models with multiple light fields, deviations from these initial statistics are typically slow-roll suppressed.}.  
After all modes have exited the horizon, one can then describe the evolution of perturbations in terms of the separate Universe picture/$\delta N$ formalism, where evolution is classical and local (in the sense discussed above). The initial adiabatic perturbations are then non-linearly conserved,
but the entropy perturbation can be transferred into $\zeta$ at both linear and non-linear order. Moreover, a purely adiabatic perturbation ($\sigma_* = 0$) remains adiabatic. In other words, evolution gives
\bea
\label{eq:multifield evol}
\zeta_* \to \zeta &=& \zeta_* + f(\sigma_*) = \zeta_* + N_{\sigma_*} \, \sigma_* + \frac{1}{2} \, N_{\sigma_*  \sigma_*} \, \sigma_*^2 + \dots \nonumber \\
\sigma_* \to \sigma &=& g(\sigma_*) \, ,
\eea
where $N_{\sigma_*}$ and $N_{\sigma_* \sigma_*}$ refer to the fact that in the $\delta N$ formalism, the effect of the initial isocurvature perturbation can be computed as the response of the number of e-foldings of expansion up to a constant-density hypersurface.
Thus, in this scenario, we end up with exactly our generalized local ansatz (\ref{eq:multifield local}), where $\za$ and $\sigma_g$ now have the physical interpretation of (Gaussian components of) the initial curvature and isocurvature perturbations at horizon exit.

We can understand from this example where our implicit assumptions are necessary.  The critical simplification is that we reduced the problem from four real solutions down to two, the growing modes $\zeta_\star$ and $\sigma_\star$.  If we set $\sigma_\star = 0$, then we are by definition in the adiabatic attractor solution and, by definition, we must reproduce all the predictions of the single-field consistency conditions.  This is what forces $\zeta|_{\sigma_\star = 0 } = \zeta_\star$.  Furthermore, having truncated the number of solutions, the second solution can always be rewritten in terms of the initial condition for the isocurvature mode, $\sigma_\star$.  If we allow for non-trivial mode coupling at horizon crossing, but retain the truncation of the superhorizon solutions, we can generate mode coupling of the from $(\zeta_S)^m \, (\sigma_L)^n$, but no coupling to $\zeta_L$ beyond those in (\ref{eq:initial NG}).  Although the consistency conditions allow mode coupling between $\zeta_S$ and $\dot \zeta_L$, the evolution requires that $\dot \zeta \propto f(\sigma_\star)$, and we can always rewrite the result in terms of the isocurvature mode.

The more dramatic modification to the local ansatz occurs when the ``decaying" modes are no longer negligible.  It remains generally true that when we set $\sigma(\x, t) = 0$, we must reproduce all the predictions of single-field inflation; yet, a more general model allows higher order mixing between $\sigma$ and $\zeta$, like those appearing in the EFT of multifield inflation~\cite{Senatore:2010wk}.  In deriving Equations~(\ref{eq:multifield local}) and (\ref{eq:multifield evol}), we were able to forbid all such terms by symmetry.  However, in doing this, we were assuming that $\za(\x,t)$ is the solution that is constant outside the horizon.  Of course, there is always a second solution that violates this assumption, but typically decays as $a^{-3}$ and plays no role in the dynamics.  However, with sufficiently rapid time dependence, sharp turns in field space, or other non-trivial dynamics, the decaying modes may not be negligible at some time during inflation and may generate non-trivial mode couplings\footnote{One may wonder how such contributions can arise without violating the symmetries in Equation~(\ref{eq:local trans}).  Because the decaying mode necessarily depends on time, one can include terms of the form $\int dt' \dot \zeta_{{\rm ad},L}(t')$ that are manifestly invariant under (\ref{eq:local trans}) but are proportional only to the decaying mode.  These terms are non-local in time in our ansatz, but are perfectly consistent with local time evolution.  This is simply a reflection that the statistics have a memory of the past evolution (which is the same reason they encode information about inflation when we measure them much later).}.  In fact, if we allow for non-attractor solutions (i.e.~the constant mode is the decaying mode), we may violate the consistency conditions even in single-field inflation~\cite{Namjoo:2012aa,Assassi:2012et,Chen:2013aj,Flauger:2013hra,mooijpalma15}.

\vskip 10pt

Now that we have covered the physical interpretation of the generalized local ansatz, let us briefly consider its implications.
Although $\za$ must always be present to maintain diffeomorphism invariance, when $P_{\za} \ll P_{\sigma}$ we can effectively neglect $\za$ for the purpose of computing statistics.  In this limit, we will reproduce the results of the standard local ansatz.  More generally, one should include both terms.  For example, if we compute $\fNL$ using Equation (\ref{eq:multifield local}) we have
\bea
\fNL &=& \frac{5}{12} \, \frac{\langle \zeta_L \, \zeta_S \, \zeta_S \rangle'}{\langle \zeta_L^2 \rangle' \, \langle \zeta_S^2 \rangle'} \nonumber \\
&=&   \frac{5}{12} \frac{\Big[ - P_{\zaL} (3 + \frac{\partial}{\partial \ln k} ) (P_{\zaS} + P_{\sigma_S}) + 4 \, c_2 P_{\sigma_L} P_{\sigma_S} \Big]}{(P_{\zaL}+P_{\sigma_L}) (P_{\zaS}+P_{\sigma_S})}  \label{eqn:fnldef}\ .
\eea
It is easy to see that the first term is a statement of the consistency conditions in the presence of $\sigma$.  Furthermore, the contribution to $\fNL$ from each term is suppressed by the relative contribution $\sigma_L$ or $\zaL$ makes to $\zeta_L$.  Now if we take the limit $P_{\sigma_L} \gg P_{{\zaL}}$ or $P_{\sigma_L} \ll P_{{\zaL}}$ we effectively return to the local ansatz or the single-field consistency conditions respectively.

For higher $N$-point functions, the presence of $\za$ and $\sigma$ with $\langle \za \, \sigma \rangle =0$ will also lead to stochastic non-Gaussianity (and scale-dependent stochastic bias~\cite{Baumann:2012bc}).  Specifically, the collapsed limits of higher $N$-point functions will be enhanced relative to the expectation from lower $N$-point functions.  For example if $c_2 \gg (n_s-1)$, $\tau_{\rm NL}$ is given by
\bea
\tau_{\rm NL} &=& \frac{1}{4}\frac{1}{P_{\zeta_L} P_{\zeta_S}^2} \langle \zeta(\k_S - \k_L) \zeta(-\k_S) \zeta(\k'_S + \k_L) \zeta(-\k'_S) \rangle' \nonumber \\
&\approx&  4 \,c_2^2  \frac{P_{\sigma_L}  P_{\sigma_S}^2 }{P_{\zeta_L} P_{\zeta_S}^2}  \nonumber \\
&\approx&  \left(\frac{6}{5}  \fNL \right)^2 \frac{(P_{\za_L} +P_{\sigma_L})}{P_{\sigma_L} } \ ,
\eea
where the last line follows from Equation~(\ref{eqn:fnldef}) and $P_{\zeta_L} \equiv P_{\za_L} +P_{\sigma_L}$. We see that the amplitude is enhanced by $\frac{(P_{\za_L} +P_{\sigma_L})}{P_{\sigma_L} } \geq 1$ relative to the expectation from local ansatz with a single field\footnote{The Suyama-Yamaguchi inequality~\cite{suyamayama08}, $\tau_{\rm NL} \geq  \left(\frac{6}{5} f_{\rm NL} \right)^2$, must always be satisfied~\cite{Smith:2011if,Assassi:2012zq} but is saturated for a single degree of freedom (up to loop corrections~\cite{Tasinato:2012js}).}, namely $\tau_{\rm NL} = \left(\frac{6}{5} f_{\rm NL} \right)^2$.  The reason is that the non-collapsed $N$-point functions are suppressed by the correlation coefficient of $\sigma$ with $\zeta$ because we do not observe $\sigma_L$ directly.  This additional suppression does not arise in collapsed configurations where we do not need to directly measure $\sigma_L$ to be sensitive to its mode coupling.  It is the same reason that one finds scale-dependent stochastic bias in these models~\cite{Baumann:2012bc}; halos are biased with respect to $\sigma_L$ which is not fully correlated with the linear density field.

\section{Deprojecting the Long Mode}
\label{sec:deprojecting}

We showed in Section \ref{sec:violation} that the single-field consistency conditions are more than just statements about the squeezed limit bispectrum, but instead dictate the response of the full short-wavelength statistics to a long mode.  Specifically, in terms of the statistically independent fluctuation, $\tilde{\zeta}_S(\x)$, it is a remapping of coordinates by the long mode,
\beq
\label{eq:cc sec4}
\zeta_S(\x) = \tilde \zeta_S( e^{\zeta_L(\x)} \x) \ .
\eeq
One way of testing this condition in all its richness is to study various $N$-point functions, correlating the long mode with powers of the short mode, e.g.~$\langle \zeta_L \, \zeta_S^n\rangle$.
An intriguing alternative follows from the realization that the remapping in Eq.~(\ref{eq:cc sec4})
is reminiscent of the effect of the gravitational lensing deflection field on cosmic microwave background (CMB) fluctuations (see e.g.~\cite{Lewis:2006fu} for review).  For example, lensing of CMB temperature is given by
\beq
T(\x) = \tilde T(\x + \boldsymbol{\alpha}(\x)),
\eeq
where $T(\x)$ is the lensed CMB temperature, $\tilde{T}$ the unlensed temperature, and $\boldsymbol{\alpha}(\x)$ is the deflection field.
In the CMB, given a measurement of the lensed temperature map $T(\x)$, it is well known that one can reconstruct
the actual {\it realization} of the lensing deflection field and then ``delens'' the CMB fluctuations to obtain $\tilde{T}$ (see e.g.~\cite{Hirata:2003ka,Smith:2010gu,Green:2016cjr}). It should therefore be possible, in principle, to do the same in the present context, i.e.~use an estimate of the long mode, $\hat{\zeta}_L$ (we will use hats to denote estimators), to locally map $\zeta_S$ back to $\tilde{\zeta}_S$, assuming the consistency conditions,
\beq\label{eq:deprojection}
\hat{\tilde{\zeta}}_S(\x) \equiv \hat{\zeta}_S(e^{-\hat{\zeta}_L} \, \x).
\eeq
Assuming $\hat{\zeta}_L$ is unbiased, the resulting ``deprojected'' short mode thus gives the fluctuations in a local unperturbed coordinate system, i.e.~the fluctuations as they would appear to a local observer\footnote{In general, one can test the consistency conditions by considering any local observable and testing if it depends on the long mode. Another good example is halo number density, which can only depend on local physics. If the consistency conditions hold, this quantity can not be modulated by $\zeta_L$ (modulo gradients of $\zeta_L$) so that the $\propto k^{-2}$ scale-dependent bias has to be exactly zero \cite{Pajer:2013ana,dePutter:2015vga,Dai:2015jaa}.}.
If the consistency conditions indeed hold, these local fluctuations should be completely independent of the long mode,
\beq
\hat{\tilde{\zeta}}_S(\x) \to \tilde{\zeta}_S(\x).
\eeq
Technically speaking, the procedure defined in Equation (\ref{eq:deprojection}) does not perfectly deproject the long mode, due to the position dependence of the long mode, but this procedure can be promoted to an exact inversion along same lines as delensing in the CMB.

Thus, one can test the consistency conditions by comparing the local statistics of the deprojected short mode in different spatial patches, and checking that they are independent of $\zeta_L$. These local statistics can be $N$-point functions of $\hat{\tilde{\zeta}}_S$ or histograms of the mode amplitudes, or another statistic. The point is that the consistency conditions predict that any local statistic will have to be independent of the long mode.

For the estimate of the long mode $\hat{\zeta}_L$, there are two scenarios. First, one could imagine measuring it directly from large-scale structure. Second, one could take the CMB lensing analogy further, and reconstruct the realization of the long mode directly from the statistics of the short modes assuming the consistency conditions.  By analogy with the quadratic estimator for lensing reconstruction, we have
\beq
\label{eq:lensingestimator2}
\hat \zeta_L^{\rm q.e.}({\bf k}) = N({\bf k}) \, \int d^3 {\bf k}' \, \zeta_S({\bf k}') \, \zeta_S({\bf k} - {\bf k}') \, g({\bf k}', {\bf k}) \, .
\eeq
If we assume that the consistency conditions hold, we can make our estimator unbiased at first order in $\zeta_L$ by requiring that
\bea
\label{eq:unbiased_requirement}
\zeta_L({\bf k}) &=& \left\langle\hat \zeta^{\rm q.e.}_L({\bf k})\right\rangle_{\zeta_S}' =  N({\bf k}) \, \int d^3 {\bf k}' \, \left\langle \zeta_S({\bf k}') \, \zeta_S({\bf k} - {\bf k}')\right\rangle' \, g({\bf k}', {\bf k}) \nonumber \\
& \approx & - N({\bf k}) \, \int d^3 {\bf k}' \, (n_s - 1) P_S(k') \zeta_L({\bf k}) \, g({\bf k}', {\bf k}) \, ,
\eea
where we have used Eq.~(\ref{eq:cc all orders1}) in the second line.  This then fixes our choice of $N({\bf k})$ to be
\beq
\label{eq:reconstruction_normalization}
N({\bf k})^{-1} = - (n_s-1) \, \int d^3 {\bf k}' \,  P_S(k') \, g({\bf k}', {\bf k}) \, .
\eeq
One could go on to define the weights $g({\bf k}', {\bf k})$ which minimize the variance of the estimator for a particular set of observations of the short modes, but that will not be necessary here.

Note, however, that if the long mode is estimated via ``lensing'' reconstruction, Eq.~(\ref{eq:lensingestimator2}), $\hat{\zeta}_L$ will be biased
if the consistency conditions are violated.  To leading order in $n_s-1$ and $\zeta_L$, we can estimate this bias by
\beq
\label{eq:reconstruction_bias}
\hat{\zeta}^{\rm q.e.}_L(\k) \approx \frac{ \int d^3 {\bf k}' \, \left\langle \zeta_S(\k') \, \zeta_S(\k- \k') \, \zeta_L(\k) \right\rangle'  P^{-1}_L(k) \, g(\k', \k)}{- (n_s - 1) \, \int d^3 k' \,  P_S(k') \, g({\bf k}', {\bf k})} \, \, \zeta_L(\k) \, .
\eeq
We see that the leading bias is determined by the squeezed limit of the three-point function.  However, if we do not have an independent measure of $\zeta_L$ we cannot see this bias directly.  Furthermore, for the local ansatz we would also find that the variance of $\hat{\tilde{\zeta}}_S(\x) = \hat{\zeta}_S(e^{-\hat{\zeta}^{\rm q.e.}_L} \, \x)$ is independent of $\zeta_L$ {\it despite} the consistency conditions being violated,
\bea
\left\langle \hat{\tilde{\zeta}}_S \hat{\tilde{\zeta}}_S \right\rangle' = 4 c_2 P_S(k) \zeta_L + (n_s-1) P_S(k) \hat \zeta^{\rm q.e.}_L + {\cal O}(\zeta_L^2) = {\cal O}(\zeta_L^2) \ . 
\eea
Since our quadratic estimator is only unbiased at linear order in $\zeta_L$ when the consistency conditions apply, we will see no visible mode coupling in the power spectrum to the expected level of accuracy.  Nevertheless, violations would show up in higher order correlation functions
\bea
\left\langle \hat{\tilde{\zeta}}_S \hat{\tilde{\zeta}}_S \hat{\tilde{\zeta}}_S \right\rangle' &=& 6 c_2P_S(k)^2 + 24 c_2^2 P_S(k)^2\zeta_L + 12 c_2 (n_s-1)P_S(k)^2 \hat \zeta^{\rm q.e.}_L + {\cal O}(\zeta_L^2) \nonumber \\
&=& 6 c_2P_S(k)^2 - 24 c_2^2 P_S(k)^2\zeta_L + {\cal O}(\zeta_L^2) \ ,
\eea
where we set $c_3 = 0$ for simplicity.  Since we are only able to check mode coupling to linear order in $\zeta_L$, this mode coupling can be made to vanish with an appropriate choice of $c_3$.  

Ultimately, the analogy with CMB lensing is limited because we want to define a procedure that works to all orders in $\zeta_L$ rather than just linear order, as defined by the quadratic estimator.  Fortunately, we can measure $\zeta_L$ directly rather than inferring it through mode coupling.  With such a measurement, one can directly check the bias of the quadratic estimator as a test of the consistency conditions.  A direct measurement of $\zeta_L$ can also be used to deproject $\zeta_S$ to all orders in $\zeta_L$ when the consistency conditions are satisfied.  If $\zeta$ is determined by the local ansatz, then we will find that for some $n,m$ with $m \geq 1$ and $n+m\leq 4$, such that $\langle \hat{\tilde{\zeta}}_S^n  \, \hat \zeta_L^m \rangle \neq 0$.  Since the consistency conditions require that $\hat{\tilde{\zeta}}_S(\x)$ is statistically independent of $\hat \zeta_L$, the presence of any non-zero contribution defines the violation of the consistency conditions when using the deprojected modes.

The description here is an idealized description of deprojection and is more challenging to implement on real observables.  In reality, we do not have the luxury of observing $\zeta(\x)$ directly, but instead see projection effects due to redshifts, lensing, recombination, etc.~\cite{Yoo:2009au,Bonvin:2011bg,Challinor:2011bk,Bruni:2011ta,Jeong:2011as,Yoo:2011zc}.  One may hope to separate the three-dimensional projections from the consistency conditions for these other projections.  Showing that this procedure can be implemented in practice is beyond the scope of this work.  From a conceptual point of view, this method of deprojection highlights that the single-field consistency conditions are a statement about about the universe for every realization of $\zeta_L$, rather than just its statistics, and can therefore be removed realization-by-realization.

\section{Discussion}
\label{sec:discussion}

Local non-Gaussianity as parametrized by the local ansatz is a natural consequence of many scenarios that convert isocurvature fluctuations into curvature perturbations at late times.  Such situations arise frequently in both multifield inflation and alternatives to inflation and is therefore a compelling target for current and future observations.  Meanwhile, single-field inflation makes a very specific set of predictions for the same correlation functions that are predicted by the local ansatz. Thus, a common way of observationally distinguishing between single-field inflation and its alternatives is by measuring local non-Gaussianity parameters.  For instance, the consistency conditions predict a squeezed limit bispectrum corresponding to $f_{\rm NL} = -\frac{5}{12} (n_s-1)$ in the local ansatz and any deviation from this points to a clear violation of single-field inflation\footnote{Violations within single-field inflation are possible by violating some of the technical assumptions discussed in Section~\ref{sec:consistency}~\cite{Namjoo:2012aa,Assassi:2012et,Chen:2013aj,Flauger:2013hra,mooijpalma15}}.

On the other hand, the local ansatz makes statements of a fundamentally different nature than the consistency conditions, and it is not a priori clear that constraining local non-Gaussianity is equivalent to testing the single-field consistency conditions.
In this article, we have attempted to clarify the relation between these two approaches.

First, we have shown that, while the local ansatz can reproduce, e.g., the single-field prediction for the squeezed limit bispectrum, it is impossible to agree with the consistency conditions to all orders,
so that the local ansatz is in general inconsistent with single-field inflation.
Thus, in principle, precision measurements of the correlation functions validating the consistency relations can rule out the local ansatz and confirm the single-field consistency conditions.  This is nontrivial in the sense that by choosing coefficients carefully, the local ansatz can match the prediction of single-field inflation for any \emph{one} correlation function.  However, we have showed that there is no choice of coefficients that may satisfy \emph{all} the conditions simultaneously.  Violations must appear which are at least of order $(n_s-1)^2$.  

Secondly, we have noted that, even in multifield inflation, a weaker version of the consistency conditions persists, namely the fact that small-scale statistics should be independent of an {\it adiabatic} shift in the long mode. This means that, technically, the usual local ansatz is inconsistent even with multifield inflation. However, the local ansatz can be generalized in a simple way, by explicitly adding a second field (loosely identified with the isocurvature fluctuation), to make it explicitly consistent with these consistency conditions. This generalized form reduces to the usual local form in the limit where the final curvature fluctuations are dominated by the second field, and reduces to the single-field prediction in the limit where the second field is negligible.

Finally, we have suggested a novel way of testing the consistency conditions. Instead of studying a hierarchy of $N$-point functions, one could follow an approach analogous to delensing of the cosmic microwave background, i.e.~remove the effect of the long mode from the short modes assuming the consistency conditions, and then check that the short-wavelength statistics are indeed independent of the long mode.

In practice, the minimal deviation of the local ansatz from the single-field consistency conditions is unobservably small.  Nevertheless, understanding the precise predictions of these models provides an important framework for future tests of inflation and its alternatives.  It is often argued that measuring $f_{\rm NL} = -\frac{5}{12} (n_s-1)$ would confirm single-field inflation.  This view has been challenged on the ground that this prediction does not require inflation but only that the short wavelength modes are statistically independent of the long wavelength modes in physical coordinates~\cite{Senatore:2012wy,Pajer:2013ana,Dai:2015jaa}. In this work, we showed that even if the mode coupling underlying this relation is ``trivial'' in physical coordinates, it can never be reproduced locally in space after inflation. As a consequence, any physical observable, such as scale-dependent bias, should therefore show a minimum violation of the consistency conditions in a universe governed by the local ansatz.

\vskip23pt
\paragraph{Acknowledgements}
We thank Daniel Baumann, Rafael Porto and Alex van Engelen for helpful discussions. D.G.~was supported by an NSERC Discovery Grant and the Canadian Institute for Advanced Research. R.d.P. and O.D. acknowledge support by the Heising-Simons foundation. J.M.~was supported by the Vincent and Beatrice Tremaine Fellowship. Part of the research described in this paper was carried out at the Jet Propulsion Laboratory, California Institute of Technology, under a contract with the National Aeronautics and Space Administration.

\clearpage
\phantomsection
\addcontentsline{toc}{section}{References}
\bibliographystyle{utphys}
\bibliography{Refs}

\providecommand{\href}[2]{#2}\begingroup\raggedright\begin{thebibliography}{10}

\bibitem{Maldacena:2015bha}
J.~Maldacena, ``{A model with cosmological Bell inequalities},''
  \href{http://dx.doi.org/10.1002/prop.201500097}{{\em Fortsch. Phys.}
  {\bfseries 64} (2016) 10--23},
\href{http://arxiv.org/abs/1508.01082}{{\ttfamily arXiv:1508.01082 [hep-th]}}.

\bibitem{Creminelli:2004yq}
P.~Creminelli and M.~Zaldarriaga, ``{Single field consistency relation for the
  3-point function},''
  \href{http://dx.doi.org/10.1088/1475-7516/2004/10/006}{{\em JCAP} {\bfseries
  0410} (2004) 006},
\href{http://arxiv.org/abs/astro-ph/0407059}{{\ttfamily arXiv:astro-ph/0407059
  [astro-ph]}}.

\bibitem{Baumann:2014cja}
D.~Baumann, D.~Green, and R.~A. Porto, ``{B-modes and the Nature of
  Inflation},'' \href{http://dx.doi.org/10.1088/1475-7516/2015/01/016}{{\em
  JCAP} {\bfseries 1501} no.~01, (2015) 016},
\href{http://arxiv.org/abs/1407.2621}{{\ttfamily arXiv:1407.2621 [hep-th]}}.

\bibitem{Baumann:2015nta}
D.~Baumann, D.~Green, H.~Lee, and R.~A. Porto, ``{Signs of Analyticity in
  Single-Field Inflation},''
  \href{http://dx.doi.org/10.1103/PhysRevD.93.023523}{{\em Phys. Rev.}
  {\bfseries D93} no.~2, (2016) 023523},
\href{http://arxiv.org/abs/1502.07304}{{\ttfamily arXiv:1502.07304 [hep-th]}}.

\bibitem{Alvarez:2014vva}
M.~Alvarez {\em et~al.}, ``{Testing Inflation with Large Scale Structure:
  Connecting Hopes with Reality},''
\href{http://arxiv.org/abs/1412.4671}{{\ttfamily arXiv:1412.4671
  [astro-ph.CO]}}.

\bibitem{Maldacena:2002vr}
J.~M. Maldacena, ``{Non-Gaussian features of primordial fluctuations in single
  field inflationary models},''
  \href{http://dx.doi.org/10.1088/1126-6708/2003/05/013}{{\em JHEP} {\bfseries
  05} (2003) 013},
\href{http://arxiv.org/abs/astro-ph/0210603}{{\ttfamily arXiv:astro-ph/0210603
  [astro-ph]}}.

\bibitem{Baldauf:2011bh}
T.~Baldauf, U.~Seljak, L.~Senatore, and M.~Zaldarriaga, ``{Galaxy Bias and
  non-Linear Structure Formation in General Relativity},''
  \href{http://dx.doi.org/10.1088/1475-7516/2011/10/031}{{\em JCAP} {\bfseries
  1110} (2011) 031},
\href{http://arxiv.org/abs/1106.5507}{{\ttfamily arXiv:1106.5507
  [astro-ph.CO]}}.

\bibitem{Creminelli:2012ed}
P.~Creminelli, J.~Nore{\~n}a, and M.~Simonovi\'c, ``{Conformal consistency
  relations for single-field inflation},''
  \href{http://dx.doi.org/10.1088/1475-7516/2012/07/052}{{\em JCAP} {\bfseries
  1207} (2012) 052},
\href{http://arxiv.org/abs/1203.4595}{{\ttfamily arXiv:1203.4595 [hep-th]}}.

\bibitem{PhysRevD.31.1792}
D.~H. Lyth, ``Large-scale energy-density perturbations and inflation,''
  \href{http://dx.doi.org/10.1103/PhysRevD.31.1792}{{\em Phys. Rev. D}
  {\bfseries 31} (Apr, 1985) 1792--1798}.
  \url{http://link.aps.org/doi/10.1103/PhysRevD.31.1792}.

\bibitem{Salopek:1990jq}
D.~S. Salopek and J.~R. Bond, ``{Nonlinear evolution of long wavelength metric
  fluctuations in inflationary models},''
\href{http://dx.doi.org/10.1103/PhysRevD.42.3936}{{\em Phys. Rev.} {\bfseries
  D42} (1990) 3936--3962}.

\bibitem{Weinberg:2003sw}
S.~Weinberg, ``{Adiabatic modes in cosmology},''
  \href{http://dx.doi.org/10.1103/PhysRevD.67.123504}{{\em Phys. Rev.}
  {\bfseries D67} (2003) 123504},
\href{http://arxiv.org/abs/astro-ph/0302326}{{\ttfamily arXiv:astro-ph/0302326
  [astro-ph]}}.

\bibitem{Creminelli:2013mca}
P.~Creminelli, J.~Nore{\~n}a, M.~Simonovi\'c, and F.~Vernizzi, ``{Single-Field
  Consistency Relations of Large Scale Structure},''
  \href{http://dx.doi.org/10.1088/1475-7516/2013/12/025}{{\em JCAP} {\bfseries
  1312} (2013) 025},
\href{http://arxiv.org/abs/1309.3557}{{\ttfamily arXiv:1309.3557
  [astro-ph.CO]}}.

\bibitem{dePutter:2015vga}
R.~de~Putter, O.~Dor\'e, and D.~Green, ``{Is There Scale-Dependent Bias in
  Single-Field Inflation?},''
  \href{http://dx.doi.org/10.1088/1475-7516/2015/10/024}{{\em JCAP} {\bfseries
  1510} no.~10, (2015) 024},
\href{http://arxiv.org/abs/1504.05935}{{\ttfamily arXiv:1504.05935
  [astro-ph.CO]}}.

\bibitem{Dai:2015jaa}
L.~Dai, E.~Pajer, and F.~Schmidt, ``{On Separate Universes},''
  \href{http://dx.doi.org/10.1088/1475-7516/2015/10/059}{{\em JCAP} {\bfseries
  1510} no.~10, (2015) 059},
\href{http://arxiv.org/abs/1504.00351}{{\ttfamily arXiv:1504.00351
  [astro-ph.CO]}}.

\bibitem{Ade:2015ava}
{\bfseries Planck} Collaboration, P.~A.~R. Ade {\em et~al.}, ``{Planck 2015
  results. XVII. Constraints on primordial non-Gaussianity},''
\href{http://arxiv.org/abs/1502.01592}{{\ttfamily arXiv:1502.01592
  [astro-ph.CO]}}.

\bibitem{RdPDore14}
R.~{de Putter} and O.~{Dor{\'e}}, ``{Designing an Inflation Galaxy Survey: how
  to measure $\sigma(f\_{\rm NL}) \sim 1$ using scale-dependent galaxy bias},''
  {\em ArXiv e-prints} (Dec., 2014) ,
  \href{http://arxiv.org/abs/1412.3854}{{\ttfamily arXiv:1412.3854}}.

\bibitem{SPHERExWP}
O.~{Dor{\'e}}, J.~{Bock}, M.~{Ashby}, P.~{Capak}, A.~{Cooray}, R.~{de Putter},
  T.~{Eifler}, N.~{Flagey}, Y.~{Gong}, S.~{Habib}, K.~{Heitmann}, C.~{Hirata},
  W.-S. {Jeong}, R.~{Katti}, P.~{Korngut}, E.~{Krause}, D.-H. {Lee},
  D.~{Masters}, P.~{Mauskopf}, G.~{Melnick}, B.~{Mennesson}, H.~{Nguyen},
  K.~{{\"O}berg}, A.~{Pullen}, A.~{Raccanelli}, R.~{Smith}, Y.-S. {Song},
  V.~{Tolls}, S.~{Unwin}, T.~{Venumadhav}, M.~{Viero}, M.~{Werner}, and
  M.~{Zemcov}, ``{Cosmology with the SPHEREX All-Sky Spectral Survey},'' {\em
  ArXiv e-prints} (Dec., 2014) ,
  \href{http://arxiv.org/abs/1412.4872}{{\ttfamily arXiv:1412.4872}}.

\bibitem{ferrsmith14}
S.~{Ferraro} and K.~M. {Smith}, ``{Using large scale structure to measure
  f$_{NL}$ , g$_{NL}$ and {$\tau$}$_{NL}$},''
  \href{http://dx.doi.org/10.1103/PhysRevD.91.043506}{{\em \prd} {\bfseries 91}
  no.~4, (Feb., 2015) 043506}, \href{http://arxiv.org/abs/1408.3126}{{\ttfamily
  arXiv:1408.3126}}.

\bibitem{yamauchietal14}
D.~{Yamauchi}, K.~{Takahashi}, and M.~{Oguri}, ``{Constraining primordial
  non-Gaussianity via a multitracer technique with surveys by Euclid and the
  Square Kilometre Array},''
  \href{http://dx.doi.org/10.1103/PhysRevD.90.083520}{{\em \prd} {\bfseries 90}
  no.~8, (Oct., 2014) 083520}, \href{http://arxiv.org/abs/1407.5453}{{\ttfamily
  arXiv:1407.5453}}.

\bibitem{lythetal03}
D.~H. {Lyth}, C.~{Ungarelli}, and D.~{Wands}, ``{Primordial density
  perturbation in the curvaton scenario},''
  \href{http://dx.doi.org/10.1103/PhysRevD.67.023503}{{\em \prd} {\bfseries 67}
  no.~2, (Jan., 2003) 023503},
  \href{http://arxiv.org/abs/astro-ph/0208055}{{\ttfamily astro-ph/0208055}}.

\bibitem{zal04}
M.~{Zaldarriaga}, ``{Non-Gaussianities in models with a varying inflaton decay
  rate},'' \href{http://dx.doi.org/10.1103/PhysRevD.69.043508}{{\em \prd}
  {\bfseries 69} no.~4, (Feb., 2004) 043508},
  \href{http://arxiv.org/abs/astro-ph/0306006}{{\ttfamily astro-ph/0306006}}.

\bibitem{Li:2008gg}
M.~Li and Y.~Wang, ``{Consistency Relations for Non-Gaussianity},''
  \href{http://dx.doi.org/10.1088/1475-7516/2008/09/018}{{\em JCAP} {\bfseries
  0809} (2008) 018},
\href{http://arxiv.org/abs/0807.3058}{{\ttfamily arXiv:0807.3058 [hep-th]}}.

\bibitem{Senatore:2012wy}
L.~Senatore and M.~Zaldarriaga, ``{A Note on the Consistency Condition of
  Primordial Fluctuations},''
  \href{http://dx.doi.org/10.1088/1475-7516/2012/08/001}{{\em JCAP} {\bfseries
  1208} (2012) 001},
\href{http://arxiv.org/abs/1203.6884}{{\ttfamily arXiv:1203.6884
  [astro-ph.CO]}}.

\bibitem{Pajer:2013ana}
E.~Pajer, F.~Schmidt, and M.~Zaldarriaga, ``{The Observed Squeezed Limit of
  Cosmological Three-Point Functions},''
  \href{http://dx.doi.org/10.1103/PhysRevD.88.083502}{{\em Phys. Rev.}
  {\bfseries D88} no.~8, (2013) 083502},
\href{http://arxiv.org/abs/1305.0824}{{\ttfamily arXiv:1305.0824
  [astro-ph.CO]}}.

\bibitem{Flauger:2013hra}
R.~Flauger, D.~Green, and R.~A. Porto, ``{On squeezed limits in single-field
  inflation. Part I},'' \href{http://dx.doi.org/10.1088/1475-7516/2013/08/032,
  10.1088/1475-7516/2013/08/032/}{{\em JCAP} {\bfseries 1308} (2013) 032},
\href{http://arxiv.org/abs/1303.1430}{{\ttfamily arXiv:1303.1430 [hep-th]}}.

\bibitem{dalaletal08}
N.~{Dalal}, O.~{Dor{\'e}}, D.~{Huterer}, and A.~{Shirokov}, ``{Imprints of
  primordial non-Gaussianities on large-scale structure: Scale-dependent bias
  and abundance of virialized objects},''
  \href{http://dx.doi.org/10.1103/PhysRevD.77.123514}{{\em \prd} {\bfseries 77}
  no.~12, (June, 2008) 123514},
  \href{http://arxiv.org/abs/0710.4560}{{\ttfamily arXiv:0710.4560}}.

\bibitem{matver08}
S.~{Matarrese} and L.~{Verde}, ``{The Effect of Primordial Non-Gaussianity on
  Halo Bias},'' \href{http://dx.doi.org/10.1086/587840}{{\em \apjl} {\bfseries
  677} (Apr., 2008) L77}, \href{http://arxiv.org/abs/0801.4826}{{\ttfamily
  arXiv:0801.4826}}.

\bibitem{slosaretal08}
A.~{Slosar}, C.~{Hirata}, U.~{Seljak}, S.~{Ho}, and N.~{Padmanabhan},
  ``{Constraints on local primordial non-Gaussianity from large scale
  structure},'' \href{http://dx.doi.org/10.1088/1475-7516/2008/08/031}{{\em
  \jcap} {\bfseries 8} (Aug., 2008) 031},
  \href{http://arxiv.org/abs/0805.3580}{{\ttfamily arXiv:0805.3580}}.

\bibitem{desjacjak10}
V.~{Desjacques} and U.~{Seljak}, ``{Primordial non-Gaussianity from the
  large-scale structure},''
  \href{http://dx.doi.org/10.1088/0264-9381/27/12/124011}{{\em Classical and
  Quantum Gravity} {\bfseries 27} no.~12, (June, 2010) 124011},
  \href{http://arxiv.org/abs/1003.5020}{{\ttfamily arXiv:1003.5020}}.

\bibitem{LopezNacir:2012rm}
D.~Lopez~Nacir, R.~A. Porto, and M.~Zaldarriaga, ``{The consistency condition
  for the three-point function in dissipative single-clock inflation},''
  \href{http://dx.doi.org/10.1088/1475-7516/2012/09/004}{{\em JCAP} {\bfseries
  1209} (2012) 004},
\href{http://arxiv.org/abs/1206.7083}{{\ttfamily arXiv:1206.7083 [hep-th]}}.

\bibitem{Goldberger:2013rsa}
W.~D. Goldberger, L.~Hui, and A.~Nicolis, ``{One-particle-irreducible
  consistency relations for cosmological perturbations},''
  \href{http://dx.doi.org/10.1103/PhysRevD.87.103520}{{\em Phys. Rev.}
  {\bfseries D87} no.~10, (2013) 103520},
\href{http://arxiv.org/abs/1303.1193}{{\ttfamily arXiv:1303.1193 [hep-th]}}.

\bibitem{Mirbabayi:2015hva}
M.~Mirbabayi and M.~Simonovi\'c, ``{Effective Theory of Squeezed Correlation
  Functions},'' \href{http://dx.doi.org/10.1088/1475-7516/2016/03/056}{{\em
  JCAP} {\bfseries 1603} no.~03, (2016) 056},
\href{http://arxiv.org/abs/1507.04755}{{\ttfamily arXiv:1507.04755 [hep-th]}}.

\bibitem{Senatore:2010wk}
L.~Senatore and M.~Zaldarriaga, ``{The Effective Field Theory of Multifield
  Inflation},'' \href{http://dx.doi.org/10.1007/JHEP04(2012)024}{{\em JHEP}
  {\bfseries 04} (2012) 024},
\href{http://arxiv.org/abs/1009.2093}{{\ttfamily arXiv:1009.2093 [hep-th]}}.

\bibitem{Namjoo:2012aa}
M.~H. Namjoo, H.~Firouzjahi, and M.~Sasaki, ``{Violation of non-Gaussianity
  consistency relation in a single field inflationary model},''
  \href{http://dx.doi.org/10.1209/0295-5075/101/39001}{{\em Europhys. Lett.}
  {\bfseries 101} (2013) 39001},
\href{http://arxiv.org/abs/1210.3692}{{\ttfamily arXiv:1210.3692
  [astro-ph.CO]}}.

\bibitem{Assassi:2012et}
V.~Assassi, D.~Baumann, and D.~Green, ``{Symmetries and Loops in Inflation},''
  \href{http://dx.doi.org/10.1007/JHEP02(2013)151}{{\em JHEP} {\bfseries 02}
  (2013) 151},
\href{http://arxiv.org/abs/1210.7792}{{\ttfamily arXiv:1210.7792 [hep-th]}}.

\bibitem{Chen:2013aj}
X.~Chen, H.~Firouzjahi, M.~H. Namjoo, and M.~Sasaki, ``{A Single Field
  Inflation Model with Large Local Non-Gaussianity},''
  \href{http://dx.doi.org/10.1209/0295-5075/102/59001}{{\em Europhys. Lett.}
  {\bfseries 102} (2013) 59001},
\href{http://arxiv.org/abs/1301.5699}{{\ttfamily arXiv:1301.5699 [hep-th]}}.

\bibitem{mooijpalma15}
S.~{Mooij} and G.~A. {Palma}, ``{Consistently violating the non-Gaussian
  consistency relation},''
  \href{http://dx.doi.org/10.1088/1475-7516/2015/11/025}{{\em \jcap} {\bfseries
  11} (Nov., 2015) 025}, \href{http://arxiv.org/abs/1502.03458}{{\ttfamily
  arXiv:1502.03458}}.

\bibitem{Baumann:2012bc}
D.~Baumann, S.~Ferraro, D.~Green, and K.~M. Smith, ``{Stochastic Bias from
  Non-Gaussian Initial Conditions},''
  \href{http://dx.doi.org/10.1088/1475-7516/2013/05/001}{{\em JCAP} {\bfseries
  1305} (2013) 001},
\href{http://arxiv.org/abs/1209.2173}{{\ttfamily arXiv:1209.2173
  [astro-ph.CO]}}.

\bibitem{suyamayama08}
T.~Suyama and M.~Yamaguchi, ``Non-gaussianity in the modulated reheating
  scenario,'' \href{http://dx.doi.org/10.1103/PhysRevD.77.023505}{{\em Phys.
  Rev. D} {\bfseries 77} (Jan, 2008) 023505}.
  \url{http://link.aps.org/doi/10.1103/PhysRevD.77.023505}.

\bibitem{Smith:2011if}
K.~M. Smith, M.~LoVerde, and M.~Zaldarriaga, ``{A universal bound on N-point
  correlations from inflation},''
  \href{http://dx.doi.org/10.1103/PhysRevLett.107.191301}{{\em Phys. Rev.
  Lett.} {\bfseries 107} (2011) 191301},
\href{http://arxiv.org/abs/1108.1805}{{\ttfamily arXiv:1108.1805
  [astro-ph.CO]}}.

\bibitem{Assassi:2012zq}
V.~Assassi, D.~Baumann, and D.~Green, ``{On Soft Limits of Inflationary
  Correlation Functions},''
  \href{http://dx.doi.org/10.1088/1475-7516/2012/11/047}{{\em JCAP} {\bfseries
  1211} (2012) 047},
\href{http://arxiv.org/abs/1204.4207}{{\ttfamily arXiv:1204.4207 [hep-th]}}.

\bibitem{Tasinato:2012js}
G.~Tasinato, C.~T. Byrnes, S.~Nurmi, and D.~Wands, ``{Loop corrections and a
  new test of inflation},''
  \href{http://dx.doi.org/10.1103/PhysRevD.87.043512}{{\em Phys. Rev.}
  {\bfseries D87} no.~4, (2013) 043512},
\href{http://arxiv.org/abs/1207.1772}{{\ttfamily arXiv:1207.1772 [hep-th]}}.

\bibitem{Lewis:2006fu}
A.~Lewis and A.~Challinor, ``{Weak gravitational lensing of the cmb},''
  \href{http://dx.doi.org/10.1016/j.physrep.2006.03.002}{{\em Phys. Rept.}
  {\bfseries 429} (2006) 1--65},
\href{http://arxiv.org/abs/astro-ph/0601594}{{\ttfamily arXiv:astro-ph/0601594
  [astro-ph]}}.

\bibitem{Hirata:2003ka}
C.~M. Hirata and U.~Seljak, ``{Reconstruction of lensing from the cosmic
  microwave background polarization},''
  \href{http://dx.doi.org/10.1103/PhysRevD.68.083002}{{\em Phys. Rev.}
  {\bfseries D68} (2003) 083002},
\href{http://arxiv.org/abs/astro-ph/0306354}{{\ttfamily arXiv:astro-ph/0306354
  [astro-ph]}}.

\bibitem{Smith:2010gu}
K.~M. Smith, D.~Hanson, M.~LoVerde, C.~M. Hirata, and O.~Zahn, ``{Delensing CMB
  Polarization with External Datasets},''
  \href{http://dx.doi.org/10.1088/1475-7516/2012/06/014}{{\em JCAP} {\bfseries
  1206} (2012) 014},
\href{http://arxiv.org/abs/1010.0048}{{\ttfamily arXiv:1010.0048
  [astro-ph.CO]}}.

\bibitem{Green:2016cjr}
D.~Green, J.~Meyers, and A.~van Engelen, ``{CMB Delensing Beyond the B
  Modes},''
\href{http://arxiv.org/abs/1609.08143}{{\ttfamily arXiv:1609.08143
  [astro-ph.CO]}}.

\bibitem{Yoo:2009au}
J.~Yoo, A.~L. Fitzpatrick, and M.~Zaldarriaga, ``{A New Perspective on Galaxy
  Clustering as a Cosmological Probe: General Relativistic Effects},''
  \href{http://dx.doi.org/10.1103/PhysRevD.80.083514}{{\em Phys. Rev.}
  {\bfseries D80} (2009) 083514},
\href{http://arxiv.org/abs/0907.0707}{{\ttfamily arXiv:0907.0707
  [astro-ph.CO]}}.

\bibitem{Bonvin:2011bg}
C.~Bonvin and R.~Durrer, ``{What galaxy surveys really measure},''
  \href{http://dx.doi.org/10.1103/PhysRevD.84.063505}{{\em Phys. Rev.}
  {\bfseries D84} (2011) 063505},
\href{http://arxiv.org/abs/1105.5280}{{\ttfamily arXiv:1105.5280
  [astro-ph.CO]}}.

\bibitem{Challinor:2011bk}
A.~Challinor and A.~Lewis, ``{The linear power spectrum of observed source
  number counts},'' \href{http://dx.doi.org/10.1103/PhysRevD.84.043516}{{\em
  Phys. Rev.} {\bfseries D84} (2011) 043516},
\href{http://arxiv.org/abs/1105.5292}{{\ttfamily arXiv:1105.5292
  [astro-ph.CO]}}.

\bibitem{Bruni:2011ta}
M.~Bruni, R.~Crittenden, K.~Koyama, R.~Maartens, C.~Pitrou, and D.~Wands,
  ``{Disentangling non-Gaussianity, bias and GR effects in the galaxy
  distribution},'' \href{http://dx.doi.org/10.1103/PhysRevD.85.041301}{{\em
  Phys. Rev.} {\bfseries D85} (2012) 041301},
\href{http://arxiv.org/abs/1106.3999}{{\ttfamily arXiv:1106.3999
  [astro-ph.CO]}}.

\bibitem{Jeong:2011as}
D.~Jeong, F.~Schmidt, and C.~M. Hirata, ``{Large-scale clustering of galaxies
  in general relativity},''
  \href{http://dx.doi.org/10.1103/PhysRevD.85.023504}{{\em Phys. Rev.}
  {\bfseries D85} (2012) 023504},
\href{http://arxiv.org/abs/1107.5427}{{\ttfamily arXiv:1107.5427
  [astro-ph.CO]}}.

\bibitem{Yoo:2011zc}
J.~Yoo, N.~Hamaus, U.~Seljak, and M.~Zaldarriaga, ``{Testing General Relativity
  on Horizon Scales and the Primordial non-Gaussianity},''
\href{http://arxiv.org/abs/1109.0998}{{\ttfamily arXiv:1109.0998
  [astro-ph.CO]}}.

\end{thebibliography}\endgroup

\end{document}